\documentclass{PoS}
\newcommand{\Slash}[1]{{\ooalign{\hfil/\hfil\crcr$#1$}}}
\usepackage{color}

\title{Possible medium effects on $\eta$-$\pi^0$ mixing and $\eta\rightarrow\pi^0\pi^+\pi^-$ decay in the asymmetric nuclear matter\footnote{KUNS-2474}}

\ShortTitle{The $\eta$ decay into $\pi^0\pi^+\pi^-$ in asymmetric nuclear matter}

\author{\speaker{Shuntaro Sakai}\\
        Department of Physics, Kyoto University,
        Kitashirakawa-Oiwakecho, Sakyo-ku, Kyoto 606-8502, Japan\\
        E-mail: \email{s.sakai@ruby.scphys.kyoto-u.ac.jp}}

\author{Teiji Kunihiro\\
        Department of Physics, Kyoto University,
        Kitashirakawa-Oiwakecho, Sakyo-ku, Kyoto 606-8502, Japan\\
        E-mail: \email{kunihiro@ruby.scphys.kyoto-u.ac.jp}}

\abstract{It is known that the decay process
$\eta\rightarrow\pi^0\pi^+\pi^-$ in free space is 
possible due to the isospin-symmetry breaking in Quantum Chromodynamics (QCD),
i.e., the small mass difference between 
$u$ and $d$ quarks: The small width is intimately related with the
mixing property between $\eta$ and $\pi^0$ mesons for which the chiral anomaly play a role.
In asymmetric nuclear matter such as heavy nuclei, isospin-symmetry breaking is
large and it is expected that the mixing property of the mesons changes
significantly and the above-mentioned decay width of the $\eta$ created
in such a medium may be enhanced.
This is an intriguing possibility to revealing a medium effect on a hadron in
nuclei. We apply the in-medium chiral perturbation theory to estimate the
modification of the $\eta$-$\pi^0$ mixing angle and the partial width of
$\eta\rightarrow\pi^0\pi^+\pi^-$ in asymmetric nuclear matter as a
function of the isospin asymmetry. 
We find that these quantities can be greatly enhanced
in neutron-rich matter, which should be
deteactable in  experiment.
}

\FullConference{XV International Conference on Hadron Spectroscopy-Hadron 2013\\
		4-8 November 2013\\
		Nara, Japan }

\usepackage{amsmath}	% required for `\align' (yatex added)
\begin{document}
\section{Introduction\label{intro}}
The theoretical account of the $\eta$ decay into three $\pi$'s is a
long-standing problem of hadron physics.
%The process comes from the isospin violation and 
It was shown by  Sutherland \cite{Sutherland1966} that 
the apparant isospin breaking in this decay process 
can not be attributed to  the electromagnetic effect,
%it has been shown in
%Ref.
and it is now understood that
%  to the $\eta$ decay vanishes at the leading order, so the process is strongly
%restricted 
the primary cause is the intrinsic isospin-symmetry breaking in QCD,
 i.e., the small mass difference of $u$ and $d$ quarks.\footnote{
This $\eta$ decay process was also discussed 
%\textbf{intensively} 
in the context  of the
U$_A(1)$ anomaly \cite{Weinberg1975} and there were various attempts to
explain the experimental data
\cite{Hudnall1974,Kogut1975,Raby1976,Kawarabayashi1980}.
}
%However, the current algebra only gives a very small value in
%comparison with the experimental data \cite{Osborn1970}.
%In the other words, the $\eta$ to 3$\pi$ process is completely
%prohibited within the the isospin symmetry.
However, the tree-level value given by the current algebra with
isospin-symmetry breaking was found too 
%It has been shown that the one loop effect of the chiral
%perturbation theory gives a significant correction to the tree result
%\cite{Gasser1985} and the recent two-loop calculation in the chiral
%perturbation theory with some assumptions gives a
%fairly good agreement with the experimental data 
small value in comparison with the experimental data \cite{Osborn1970}.
The incorporation of the one loop effect in the chiral perturbation
theory leads to a significant correction to the tree result
\cite{Gasser1985} and the recent two-loop calculation in the chiral
perturbation theory with some assumptions gives a fairly good agreement
with the experimental data \cite{Bijnens2007}.
It should be stressed that the all above calculation show that the
essential cause of the $\eta$-3$\pi$ is the isospin asymmetry in QCD.
%\textbf{One of the interesting subjects of hadron physics is the change of the
%chiral properties in the environment, for example, heat bath, nuclear
%matter, electromagnetic field, and so on.}
%This suggest the possibility that the existence of the environment,
%i.e. heat bath, nuclear matter, magnetic field, etc., affect the hadron
%properties \cite{HK94}, for example, meson decay constant
%\cite{Weise2000,Jido2008}.
%The recent theoretical and experimental works investigate the partial
%restoration of chiral symmetry \cite{Suzuki2004,Friedman2004}.
%The analysis of the experimental data of the $\pi N$ system suggests the
%change of the interaction strength between $\pi$ and $N$.

%In this work, we explore a possible enhancement of the
%$\eta$-to-3$\pi$ decay in the asymmetric nuclear matter where the
%isospin-symmetry breaking is
% broken
%\textbf{enhanced}.
In this work, we examine the possible enhancement of the
$\eta$-to-3$\pi$ decay in the iso-asymmetic nuclear matter where the
isospin symmetry is explicitly broken, which may enhance the $\eta$
decay!
This can be another example of the hot subjects to explore
possible modification of hadron properties in the environment
characterized by temperature, baryonic density, magnetic field and so
on\cite{HK94}.
We estimate a change of the $\eta$-$\pi^0$ mixing angle and the
partial width of the $\eta$ decay into $\pi^0\pi^+\pi^-$ using the
in-medium chiral perturbation theory \cite{Meissner2002,Kaiser2002} at
the nucleon one loop level.
We find that the $\eta$ decay into 3$\pi$ is enhanced in the asymmetric
nuclear matter,
%\textbf{compared to the symmetric nuclear matter}
which is a novel medium effect on  hadron properties and 
can be tested in experiment.
\section{Method}
%In 
Our calculation is based on 
%, we use 
the in-medium chiral perturbation theory
\cite{Meissner2002,Kaiser2002},
% to take the asymmetric nuclear matter effect into account.
and the Lagrangian of $\pi N$ system is given as follows,
\begin{eqnarray}
 \mathcal{L}&=&\mathcal{L}_{\pi\pi}^{(2)}+\mathcal{L}_{\pi
  N}^{(1)}+\mathcal{L}_{\pi N}^{(2)},\label{lag}\\
 \mathcal{L}_{\pi \pi}^{(2)}&=&\frac{f^2}{4}{\rm tr}\partial_\mu
  U\partial^\mu U+\frac{f^2}{4}{\rm tr}(\chi U^\dagger+U\chi^\dagger), \\
 \mathcal{L}_{\pi
  N}^{(1)}&=&\bar{N}\left(i\Slash{D}-m_N+\frac{g_A}{2}\gamma^\mu\gamma_5u_\mu\right)N, \\
 \mathcal{L}_{\pi
  N}^{(2)}&=&c_1\left<\chi_+\right>-\frac{c_2}{4m_N^2}\left<u_\mu
		 u_\nu\right>\left(\bar{N}D^\mu D^\nu N+{\rm
		 h.c.}\right)+\frac{c_3}{2}\left<u_\mu
		 u^\mu\right>\bar{N}N\nonumber\\
&&-\frac{c_4}{4}\bar{N}\gamma^\mu\gamma^\nu\left[u_\mu,u_\nu\right]N+c_5\bar{N}\left[\chi_+-\frac{1}{2}\left<\chi_+\right>\right]N.
\end{eqnarray}
The nuclear matter effect is taken into account through the nucleon propagator,
\begin{eqnarray}
 (\Slash{p}+m_N)\left\{\frac{i}{p^2-m_N+i\epsilon}-2\pi\delta(p^2-m_N^2)\theta(p
_0)\theta(k_f-|\vec{p}|)\right\},\label{prop}
\end{eqnarray}
which consists of the free-space part and the Pauli blocking effect in
the nuclear medium.
In the present work, the nuclear matter is treated as a free Fermi gas
as the leading-order approximation.
The loop calculations involve the Fermi momentum
$k_f^{(p,n)}$ of the proton and the neutron; we only take into
consideration the terms of the order of the $k_f^{(p,n)3}$ assuming the
proton and neutron density $\rho_{p,n}=\frac{1}{3\pi^2}k_f^{(p,n)3}$ are small.
%The proton $\rho_{p}$ and neutron  density $\rho_{n} $ region;
% $\rho_{p,n}=\frac{1}{3\pi^2}k_f^{(p,n)3}$.
%The interaction between NG boson and nucleon is introduced through the chiral
%Lagrangian Eq.(\ref{lag}) perturbatively.
The values of the low energy constants appearing in the Lagrangian
Eq. (\ref{lag}) are taken from
Refs. \cite{Gasser1985b,Bernard1997}, where the meson masses and
decay constant are physical ones.

\begin{figure}
\centering
 \includegraphics[width=14cm]{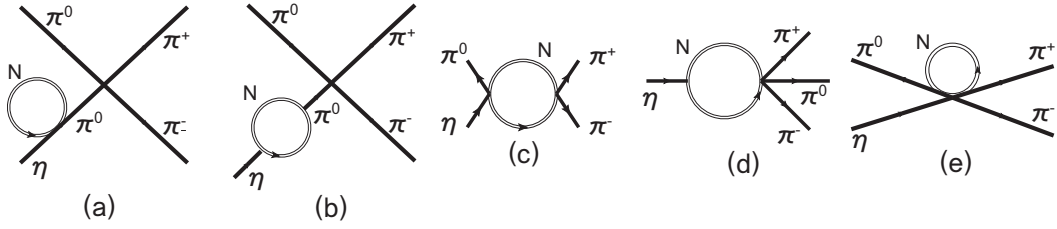}
\caption{The diagrams calculated in our calculation. The solid line
 represents the meson propagation and the double-solid line represents
 the propagation of the nucleon. The meson-baryon vertex may come from
 either $\mathcal{L}_{\pi N}^{(1)}$ or $\mathcal{L}_{\pi N}^{(2)}$}
\label{diagrams}
\end{figure}

We evaluate the contribution from the diagrams shown in
Fig. \ref{diagrams}.
In our calculation, we consider the nucleon mass $m_N$ as a large
quantity and ignore the other quantities such as the meson mass or the
three-momneta of hadrons in comparison with the nucleon mass.
Then the nucleon one-loop diagram gives contribution in the order of
$O(k_f^3)$, and the tree level in the free-space part is in the order of
$O(q^2)$.
The first two diagrams (a) and (b) cause a change in the $\eta$-$\pi^0$
mixing angle and the other diagrams (c), (d) and (e) modify the
$\eta$-to-3$\pi$ decay rate.
%\textbf{The folowing english should be elaborated.}
%We evaluate the contributions from the diagrams shown in
%Fig. \ref{diagrams}.
%We evaluate the effect of the nuclear matter on the
%$\eta$ decay process.
%Here, we consider the the nucleon one-loop order in the medium,
%$O(k_f^3)$, and the tree level in the free-space part,
%$O(q^2)$.
%The first two diagrams (a) and (b) cause a change in the
%$\eta$-$\pi^0$ mixing angle and the other diagrams (c) and (d) modify
%contribute
% the $\eta$-to-3$\pi$ decay rate.
% directly.
%These diagrams give the $O(k_f^3)$ corrections.
%contributions.
%In the present work, the nuclear matter is treated
%approximated 
%as a free Fermi gas in the leading order.
% so the Fermi momentum $k_f^{(p,n)}$ (the
%superscript $p,n$ means the Fermi momentum of the proton and  neutron,
%respectively) and 
%In our calculation, we consider the nucleon mass $m_N$ as a large quantity and
%ignore the other quantities such as the meson mass or the three-momenta of
%hadrons in comparison with the nucleon mass. 

\section{Results}
Let the $m_{\eta\pi^0}^2$ be the off-diagonal term of the $\eta$ and
$\pi^0$ meson mass matrix.
%\textbf{The $\eta$ and $\pi^0$ field in the physical basis and flavor basis are related with
%\begin{eqnarray}
%\eta=\eta_8\cos\theta-\pi_3\sin\theta,\\
%\pi^0=\eta_8\sin\theta+\pi_3\cos\theta.
%\end{eqnarray}
%Here, $\eta$ and $\pi^0$ are meson fields in the physical basis and the
%$\pi_3$ and $\eta_8$ are in the flavor basis.}
Then the  $\eta$-$\pi^0$ mixing angle $\theta$ is given as $\tan 2\theta=\frac{2m_{\eta\pi^0}^2}{m_\eta^2-m_{\pi^0}^2}$.
%The in-medium $\eta$-$\pi^0$ mixing angle $\theta$
In the medium, $m_{\eta\pi^0}^2$ is evaluated to be
\begin{eqnarray}
%\tan2\theta=\frac{2}{m_\eta^2-m_{\pi^0}^2}\left(
m_{\eta\pi^0}^2=\frac{m_1^2}{\sqrt{3}}+\frac{g_A^2m_\eta^2}{4\sqrt{3}f^2m_N}\delta\rho+\frac{4c_1m_1^2}{\sqrt{3}f^2}\rho+\frac{2c_5}{\sqrt{3}f^2}m_{\pi^0}^2 \delta\rho,
%\right). 
\label{maeq}
\end{eqnarray}
with $m_1^2=m_{K^0}^2-m_{K^\pm}^2-m_{\pi^0}^2+m_{\pi^\pm}^2$, which is 
finite
%the contribution from 
due to the current-quark mass difference between $u$ and $d$ quarks.
The $\eta$-$\pi^0$ mixing angle is small and we have 
\begin{eqnarray}
% \sin\theta\sim
\theta\sim\frac{1}{2}\tan
  2\theta=\frac{m_{\pi^0\eta}^2}{m_\eta^2-m_{\pi^0}^2}.\label{theta}
\end{eqnarray}
Here, we note that the in-medium $\eta$-$\pi^0$ mixing angle depends not
only on the nuclear asymmetry $\delta\rho=\rho_n-\rho_p$ but also on
the total density $\rho=\rho_n+\rho_p$.

\begin{figure}
\begin{minipage}[t]{0.45\hsize}
  \includegraphics[width=7cm]{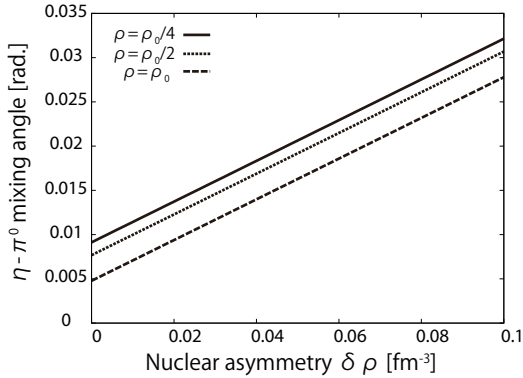}
\caption{%The in-medium $\eta$-$\pi^0$ mixing angle. 
%The horizontal and vertical  axes
% represent the proton density $\rho_p$, the neutron density $\rho_n$, respectively.
% andthe in-medium $\eta$-$\pi^0$ mixing angle, respectively. 
%The diagonal line 
%of the bottom plane 
%corresponds to the symmetric nuclear matter
% $\rho_p=\rho_n$. 
%\textbf{The following english should be elaborated:}
%The $\eta$-$\pi^0$ mixing angle is small in the blue
% region of the bottom plane and the mixing angle is large in the red region.
%\textbf{In the red and region, the $\eta$-$\pi^0$ mixing angle is large and small, respectively.}
The nuclear asymmetry $\delta\rho$ dependence of the
 $\eta$-$\pi^0$ mixing angle with fixed nuclear density $\rho$.
The holizontal axis represents the nuclear asymmetry and the vertical
 axis the mixing angle in radian.
The solid line, dotted, and dashed lines represent the mixing angle in
 the nuclear matter at $\rho=\rho_0/4$, $\rho=\rho_0/2$, and  $\rho=\rho_0$, respectively.}
\label{mafig}
\end{minipage}
\begin{minipage}[t]{0.1\hsize}
\end{minipage}
\begin{minipage}[t]{0.45\hsize}
  \includegraphics[width=7cm]{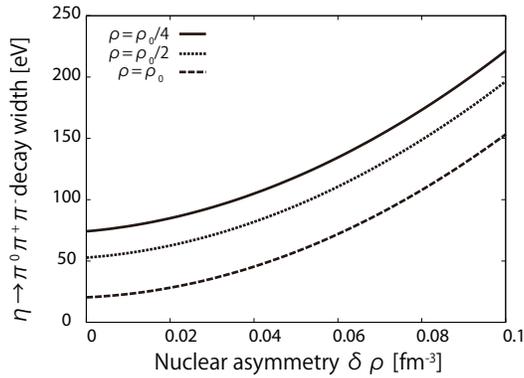}
\caption{%The in-medium decay width of $\eta\rightarrow \pi^0\pi^+\pi^-$.
%The horizontal and vertical  axes
% The $x$, $y$ axes represent 
%are the same  the left figure.
% and the $z$ axis represents the partial width of $\eta\rightarrow \pi^0\pi^+\pi^-$. 
%The decay width is small in the blue region and the decay width is large in
% the red region.} 
The nuclear asymmetry dependence of the $\eta\rightarrow
 \pi^0\pi^+\pi^-$ decay width.
The holizontal and the vertical axes represent the nuclear asymmetry and
 the $\eta\rightarrow\pi^0\pi^+\pi^-$ decay width.
The solid, dashed, and dotted lines are same as those in Fig. 2. 
}
\label{widthfig}
\end{minipage}
\end{figure}

%The $\eta$-$\pi^0$ mixing angle in the various nuclear density $\rho$ and
%the nuclear asymmetry $\delta\rho$ is shown in Fig. \ref{mafig}:
The nuclear asymmetry $\delta\rho$ dependence of the $\eta$-$\pi^0$
mixing angle with several fixed total density $\rho$ is shown in
Fig. \ref{mafig}.
We see that  the $\eta$-$\pi^0$ mixing angle
%Figure\ref{mafig} shows that 
increases in the asymmetric nuclear matter.
%\textbf{For example}, the neutron-rich matter (large $\rho_n$ with
%small $\rho_p$
%region of Fig. \ref{mafig}) enhances the $\eta$-$\pi^0$ mixing angle.
This result is natural and  confirms our
%is consistent with the 
expectation that the asymmetric
nuclear matter works to enhance the isospin-symmetry breaking effect
% and the enhancement of the isospin asymmetry 
and enlarges the $\eta$-$\pi^0$ mixing angle.

% as mentioned in Sect. \ref{intro}.

Next, the in-medium decay width of $\eta$ to $\pi^0\pi^+\pi^-$
$\mathcal{M}_{\eta\rightarrow\pi^0\pi^+\pi^-}$ is evaluated to be
\begin{eqnarray}
 \mathcal{M}_{\eta\rightarrow\pi^0\pi^+\pi^-}=\frac{\sin
  \theta}{3f^2}\left((m_\eta^2-m_{\pi^0}^2)+3(s-s_0)\right)-\frac{g_A^2}{3\sqrt{3}f^4m_N}m_\eta E_0\delta\rho.\label{widtheq}
\end{eqnarray}
Here the $\sin\theta \sim \theta$ is obtained from Eqs. (\ref{theta})
and (\ref{maeq}), $s=(p_\eta-p_{\pi^0})^2$ denotes a Mandelstam valuable with
$p_\eta$ and$p_{\pi^0}$ being the four-momenta of the  $\eta$ and
$\pi^0$, respectively, and $s_0$ is given by
%\begin{eqnarray}
$s_0=m_{\pi}^2+\frac{m_\eta^2}{3}$.
%\end{eqnarray}

%The plot of 
The $\delta\rho$ dependence of
the decay width of $\eta\rightarrow\pi^0\pi^+\pi^-$ with several fixed
$\rho$ are presented in Fig. \ref{widthfig}, which shows that the decay width
%of $\eta$ to $\pi^0\pi^+\pi^-$ 
is enhanced in the neutron-rich matter along with  the $\eta$-$\pi^0$
mixing angle.
This enhancement is traced back to the first term of the matrix element 
%of the $\eta\rightarrow\pi^0\pi^+\pi^-$shown 
in Eq. (\ref{widtheq}), which is proportional to $\sin\theta$ with
$\theta$ being the $\eta$-$\pi^0$ mixing angle.
%\textbf{($\sin^2\theta$に比例する以外の他の項も現れるから)}
%; note that the width itself is proportional 
%to the square of the mixing angle.

%\textbf{Especially, when we have compared the mixing angle and
%$\eta\rightarrow\pi^0\pi^+\pi^-$ decay width in the free space and in the
%asymmetric nuclear matter $\rho=0.06$ fm$^{-3}$ and
%$\delta\rho=0.01$ fm$^{-3}$ as an example at
%the surface of the heavy nucleus, we have found that the difference of
%the mixing angle has been about 0.15 $^\circ$ and the difference of the
%decay width has been about 50 eV.}
\section{Conclusion}
In this paper, we have examined the effect of the isospin-asymmetry of the  nuclear
medium on the $\eta$-$\pi^0$ mixing angle and the $\eta\rightarrow
\pi^0\pi^+\pi^-$ decay width.
%We have included 
The nuclear matter effect is included within the nucleon one-loop
level, $O(k_f^3)$, whereas the free-space part is evaluated in the tree
level up to $O(q^2)$.
%From the calculation, 
The resultant  $\eta$-$\pi^0$ mixing angle and
the 
%partial 
width of
$\eta\rightarrow\pi^0\pi^+\pi^-$ are presented in
Eq. (\ref{maeq}) and   Eq. (\ref{widtheq}), respectively.
We have found that
 both the 
%$\eta$-$\pi^0$
mixing angle and the 
%$\eta\rightarrow\pi^0\pi^+\pi^-$ 
decay width are largely enhanced in  neutron-rich asymmetric nuclear matter.
%\textbf{以下は本文で書いておくべき}
%\sout{As an example, we have showed the difference of the mixing angle and
%decay width in free space and the asymmetric nuclear matter,
%$\rho=0.06$ fm$^{-3}$ and $\delta\rho=0.01$ fm$^{-3}$.
%The difference of the $\eta$-$\pi^0$ mixing angle has been obtained as
%about 0.15$^\circ$ and the difference of the decay width has been
%obtained as about 50 eV.}

For the discussion of the observability of the enhancements of the
mixing angle or the decay width, we are plannning to include the higher
order contribution of meson loop, which gives the large correction to
the decay width in the case of the free space.
The large correction from the meson one loop is related to the $\sigma$
resonance in the S-wave $\pi^+\pi^-$ channel
\cite{Roiesnel1981,Gasser1985} and the further effect of 
the nuclear matter on the decay process may come from the sigma mode in
the nuclear matter.

\acknowledgments
S.~S. is a JSPS fellow and appreciates the support by a JSPS
Grant-in-Aid(No. 25-1879). 
T.K. was partially supported by a
Grant-in-Aid for Scientific Research from the Ministry of Education,
Culture, Sports, Science and Technology (MEXT) of Japan
(Nos. 20540265 and 23340067),
by the Yukawa International Program for Quark-Hadron Sciences.

\end{document}